# Attenuation characteristics of spin pumping signal due to travelling spin waves

Sankha Subhra Mukherjee, Praveen Deorani, Jae Hyun Kwon, and Hyunsoo Yang[*]

*Department of Electrical and Computer Engineering, National University of Singapore, 117576, Singapore*

The authors have investigated the contribution of the surface spin waves to spin pumping. A Pt/NiFe bilayer has been used for measuring spin waves and spin pumping signals simultaneously. The theoretical framework of spin pumping resulting from ferromagnetic resonance has been extended to incorporate spin pumping due to spin waves. Equations for the effective area of spin pumping due to spin waves have been derived. The amplitude of the spin pumping signal resulting from travelling waves is shown to decrease more rapidly with precession frequency than that resulting from standing waves and show good agreement with the experimental data.

## I. INTRODUCTION

Berger postulated the existence of spin pumping[1,2] as the inverse effect of spin torque. Its effects on Gilbert damping has been well-established[3-6] and its effects on spin valves have also been studied[7]. A dc voltage in magnetic/nonmagnetic trilayers resulting from magnetization dynamics had been measured and was attributed to spin pumping[8]. After the demonstration of the detection of a dc signal resulting from spin pumping and the inverse spin Hall effect (ISHE)[9-11], spin pumping has gained significant interest in the scientific community. Many papers have been devoted to understanding quantitative measures of various phenomenological spin pumping parameters[12-19], including the quantification of spin pumping resulting from different types of materials[20,21], and to the separation of a dc signal arising from anisotropic magnetoresistance (AMR)[18,22]. Spin pumping resulting from nonlinear magnetization dynamics has also been studied[23]. Some novel applications of spin pumping have been recently demonstrated. For example, spin pumping has been shown to be an efficient method of injecting spins into semiconductors[24], thus avoiding the impedance mismatch problem, and for transmitting electrical signals through electrical insulators[25].

Although Ferromagnetic resonance (FMR) has been nearly exclusively used in spin pumping studies, spin wave dynamics can add significant changes in the physics to the spin pumping phenomenon[26,27]. The measured signals in spin pumping phenomenon are generally relatively small, and are of the order of several microvolts. Bailleul *et al.*[28] has shown that spin waves travel significant distances over NiFe. In his measurements, he has successfully measured spin wave signals as far away as 100 μm for the generating source. In this paper, we explore the possibility of using magnetostatic spin waves (MSSW) rather than FMR for increasing the spin pumping signal levels. Furthermore, although a significant amount of theory currently exists that deals with the quantification of the spin pumping signal resulting from FMR, there is little theory which deals with the quantification of spin pumping signals resulting from MSSW. In this paper, we build upon the theoretical framework of spin pumping due to FMR and extend it to MSSW.

## II. SAMPLE FABRICATION AND MEASUREMENT SETUP

Figure 1(a) shows a schematic representation of the device. The schematic cross section of the device is shown in Fig. 1(b). All patterning is done using photo lithography and lift-off processing. First, an 800 μm × 600 μm × 14 nm Pt strip is patterned. Subsequently, 20 nm $Ni_{81}Fe_{19}$ (Py) is sputter-deposited and patterned into a 570 μm × 600 μm strip, after ion-milling 4 nm into the Pt layer to provide good electrical contact with the Pt. Then, a 640 μm × 630 μm × 30 nm $SiO_2$ strip was sputter deposited to encapsulate the Py. Finally, Ta(5 nm)|Cu(180 nm) asymmetric coplanar strips (ACPS) and the dc probe pads are patterned and sputter-deposited simultaneously. Again, before deposition, a 4 nm Ar ion-milling of Pt is performed for providing good electrical contact. The signal line width is 60 μm, the ground line width is 180 μm, and the signal-ground spacing is 30 μm. The distance between the signal lines of the ACPS is 20 μm. In Fig. 1(b), the direction of the *y*- and *z*-axes of a right-handed coordinate system is shown for describing the mathematics that follow. The bias field ($H_b$) is applied along the *x*-axis. A signal generator (SG) is used for applying a 15 dBm sinusoidal signal to one ACPS waveguide. The resultant magnetization dynamics causes spins to be pumped into the Pt layer, and due to ISHE, are subsequently converted into a charge current, and can be detected at the two ends of the Pt as a dc voltage. Simultaneously, the inductive signal at the other waveguide may be measured by a spectrum analyzer (SA), after amplification by a 29 dB low-noise amplifier.

## III. MEASUREMENT AND ANALYSIS OF MSSW

As mentioned before, the spin pumping voltage and the spin wave power output has been simultaneously measured. Spin waves generated at the excitation ACPS line by a signal generator are measured at the detection ACPS line using a spectrum analyzer. The measurement procedure for MSSW is first described, and will subsequently be analyzed.

The following procedure is used for the spin wave measurements. First, a bias field greater than 0.2 T is applied to the sample. For this bias field, the spin wave precessional frequency is greater than 13 GHz. Since spectrum analyzer measurements are carried out between 2 GHz and 6 GHz, the measured response at this value of bias field is that which results from microwave power coupled directly between the waveguides through the air and the substrate. This is the background signal. Subsequently, measurements are done at the bias fields shown in Fig. 2. The signal inductively coupled due to spin waves is extracted by subtracting the background signal from each measurement. The measured MSSW power spectrum is shown in Fig. 2 as a function of the precessional frequency ($f$), for several values of $H_b$.

The LLG equation $\partial_t \mathbf{M} = \gamma \mu_0 (\mathbf{M} \times \mathbf{H}) + (\alpha / M_S)(\mathbf{M} \times \partial_t \mathbf{M})$ is solved for the system shown in Fig. 1(b) (with the coordinate axes shown) with shape anisotropies ($N_x$, $N_y$, $N_z$), where $N_x + N_y + N_z = 1$. Because of thin-film geometry, we assume that $N_x = N_z = 0$, and $N_y = 1$. The dc magnetic bias $\mathbf{H_b} = H_b \mathbf{a_x}$. The rf field is modeled as $\mathbf{h} = h_{rf} e^{j\omega t} \mathbf{a_z}$, and has no component in the $\mathbf{a_x}$ direction. The total time-dependent magnetization may be approximated as $\mathbf{M} = M_S \mathbf{a_x} + \mathbf{m}$, where $\mathbf{m}$ is the time-dependent 'small-signal' magnetization and is assumed to be much smaller than $M_s$ in the linear regime. Furthermore, for the same reason, the magnitude of $\mathbf{M}$ is given by $|M_S \mathbf{a_x} + \mathbf{m}| = M_s$, and is independent of time in the linear regime. When this system is linearized, the magnetization dynamics may be calculated as: $\mathbf{m} = \overline{\chi} \mathbf{h}$, where the susceptibility tensor $\overline{\chi}$ is given by:

$$\bar{\bar{\chi}} = \frac{\omega_M}{D_n} \begin{bmatrix} D_n/\omega_n & 0 & 0 \\ 0 & \omega_0 - j\omega\alpha & -j\omega \\ 0 & j\omega & \omega_M + \omega_0 - j\omega\alpha \end{bmatrix} \qquad (1)$$

where $\omega_M = \gamma\mu_0 M_s$, $\omega_0 = \gamma\mu_0 H_b$, $D_n = [\omega_0 - j\omega\alpha][\omega_M - \omega_0 - j\omega\alpha] - \omega^2$, $\gamma$ is the gyromagnetic ratio of an electron, calculated with a g-factor of 2.1 for electrons in NiFe, and $\mu_0$ is the permeability of free space. $M_s$ is the saturation magnetization, and for NiFe, such that $\mu_0 M_s = 1$ T. $h_{rf}$ is calculated from Ampere's law from the equation $2L_{sg} h_{rf} = \sqrt{P_{inp}/R_{wg}}$.[29] Here, $L_{sg}$ is the width of the signal line (60 μm), $P_{inp}$ is the input power of 15 dBm (i.e. 31.62 mW), and $R_{wg}$ is the resistance of the waveguide (4.4 Ω) measured using a 4-probe technique. Thus, $h_{rf}$ is calculated as $1.608\times10^{-6}$ A/m. Note that the rf field under the ground line is a third of that under the signal line, because the ground line is three times wider.

The measurement corresponds to the real part $\text{Re}[\mathbf{m}\cdot\mathbf{a}_z]$ of the dynamic magnetization, and resembles the real part of signals routinely obtained from similar measurements using a vector network analyzer (VNA)[30]. $A_{SW}\text{Re}[\mathbf{m}\cdot\mathbf{a}_z]$ has been used to fit the measured data by solid lines in Fig. 2 for the different values of $H_b$, as previously done[30] using the fitting parameter $A_{SW}$. The value of $A_{SW}$ is found to be -0.0033. Note that this value is negative because, the orientation of the excitation and the detection spin waves are anti-symmetric. The calculations fit the measured values quite well.

**IV. MEASUREMENT AND ANALYSIS OF SPIN PUMPING**

Description of the measurement procedure and analysis of the spin pumping is provided in this section. Figure 3(a) shows the measured values of the spin pumping signal (by open squares) as a function of the precessional frequency $f$, measured simultaneously to the measurement of the spin wave signals shown in Fig. 2. During the measurement, while the sinusoidal signal generator sweeps the input frequency and the spectrum analyzer measures the resultant output MSSW power, a voltmeter connected at the two ends of the Pt strip measures the dc voltage resulting from spin pumping and ISHE. Solid lines in Fig. 3(a) correspond to calculated spin pumping voltage. The calculations for the spin pumping voltage are described below.

### A. Spin current generation

When an rf frequency **h** is applied at one of the ACPS lines, magnetization dynamics **M**(*t*) is set up within the Py. The magnetization dynamics causes a spin current $j_s = \hbar \text{Re}(2g_{\uparrow\downarrow})[\mathbf{M} \times \partial_t \mathbf{M}]/M_s^2 8\pi$ to be pumped into the adjacent Pt layer at the Pt-Py interface. Here, $g_{\uparrow\downarrow}$ represents the spin mixing conductance between Py and Pt, and has a value of ~ $2.1 \times 10^{19}$ m$^{-2}$.[4, 31] Under sinusoidal oscillations, the total dc component of the spin current over the thickness of the Pt layer may be calculated as[12, 31, 32]:

$$\mathbf{j}_s^{DC} = [\mathbf{M} \times \partial_t \mathbf{M}]_{DC} \frac{\hbar \text{Re}(g_{\downarrow\uparrow})}{4\pi M_s^2} \left[ \frac{\lambda_{sd}}{t_{Pt}} \tanh\left(\frac{t_{Pt}}{2\lambda_{sd}}\right) \right]. \tag{2}$$

where, $t_{Pt}$ is the thickness of the Pt (10 nm), and $\lambda_{sd}$ is the spin diffusion length of Pt (10 nm) at room temperature[33]. Rather than calculating the cone angle, the calculation of the precession is done directly. Note that $[\mathbf{M} \times \partial_t \mathbf{M}]_{DC} = M_s \mathbf{a_x} \times j\omega m_z \mathbf{a_z} = -j\omega m_z M_s \mathbf{a_y}$. Since $m_z$ is complex, as mentioned in Section III, measured spin wave magnetization dynamics corresponds to the real part $\text{Re}[m_z]$. Thus, the spin current density may be written as:

$$\mathbf{j}_s^{DC} = -\frac{\hbar \operatorname{Re}(g_{\downarrow\uparrow})\operatorname{Re}([j\omega m_z]_{DC})}{4\pi M_s}\left[\frac{\lambda_{sd}}{t_{Pt}}\tanh\left(\frac{t_{Pt}}{2\lambda_{sd}}\right)\right]\mathbf{a}_y \qquad (3)$$

The conversion of spin current density to spin current is phenomenologically easy, and is given simply by $\mathbf{i}_s^{DC} = WL_{tot}\mathbf{j}_s^{DC}$. Here, $W$ is the width of the Pt strip, while $L_{tot}$ is the total length over which spin pumping occurs. In the case of FMR, this value ($L_{tot}$) is simply equal to the length of the Py layer. However, in case of spin waves this value is much more difficult to calculate and the length is separated out into two parts ($L_{tot} = L_{sw}+L_{wg}$). $L_{sw}$ is a representative length responsible for a contribution to the spin pumping voltage due to travelling spin waves, and $L_{wg}$ is the length of the Py under the signal line and the ground line. While $L_{wg}$ is constant, $L_{sw}$ is not, and is a function of the applied bias field and magnetization as will be described in Section IV.C. The contribution to spin pumping due to the section of Py of length $L_{sw}$ forms one of the main aims of this study.

Magnetization dynamics in the section of Py that lies just under the waveguide results in FMR. The ground line is three times the size of the signal line, therefore $h_{rf}$ under the signal line is three times as small. Hence, the total contribution to the spin current due to the FMR under the signal line and the ground lines are the same, and as a result, $L_{wg}$ can be approximated as $L_{wg} = 2L_{sg}$, where $L_{sg}$ is the width of the signal line. Thus, the conversion from spin current density, to spin current is given by:

$$\mathbf{i}_s^{DC} = W\left(L_{sw} + 2L_{sg}\right)\mathbf{j}_s^{DC}. \qquad (4)$$

**B. Generation of voltage due to ISHE**

Because of ISHE, this spin current is converted into charge current $\mathbf{i}_c = \gamma_{sh}(2e/\hbar)\left[\mathbf{i}_s^{DC} \times \boldsymbol{\sigma}\right]$, and is detected as a voltage across the Py strip. Here $\gamma_{sh}$ represents the spin Hall angle, and has a value between 0.0067 and 0.37 for Pt at room temperature.[31, 34] Since there is little agreement between the values of the spin Hall angle, we have used 0.37 for these calculations. We use a slightly modified notation for the spin Hall angle to distinguish it from the gyromagnetic ratio used in Eq. (1). Finally, the resistance of the Pt|Py layer can be used for converting the current directly into the measured voltage.

$$V_{SP} = R_{Pt\|Py} \frac{\gamma_{sh} e}{\hbar} W \left(L_{sw} + 2L_{sg}\right) \mathbf{j}_s^{DC} \quad (5)$$

Rather than using the values of the resistivities for $R_{Pt\|Py}$, in the present experimental configuration, the resistance has been directly measured using the standard 4-probe technique, as a function of the bias field. This allows us to measure the total resistance directly and consequently eliminates unnecessary errors in the calculation. $R_{Pt\|Py}$ is measured to be 9.35 Ω, and it varies by less than 5 mΩ over the applied magnetic fields.

### C. The calculation of $L_{sw}$

The influence of spin waves on the measured spin pumping voltage is primarily governed by the length $L_{sw}$, as shown in Eq. (5). The effectiveness of the surface wave excitations may be calculated by identifying the losses of magnetization dynamics.[35] The spin wave relaxation time is defined as a time required for the amplitude of the magnetization dynamics to decay by a factor of $e^{-1}$. In Eq. (1), it is important to note that the effect of the Gilbert term in the LLG equation is to simply convert $\omega_0$ to $\omega_0 - j\omega\alpha$, (i.e. to $\omega_0 - \Delta\omega_0$). The relaxation time for such an excitations is obtained from the imaginary part of $\omega_0$ as $T_{FMR} = 1/\alpha\omega$. The dispersion relationship for surface waves is given by:

$$\omega_{MSSW}^2 = \omega_0(\omega_0 + \omega_M) + \omega_M^2[1 - \exp(-2kd)]/4 \tag{6}$$

where $k$ is the value of the wave vector of the propagating spin wave, and may be calculated as $k = 2\pi/L_{sg}$.[28, 36, 37] Normally, the wavelength is taken as the width of the signal line, and $d$ is the thickness of the Py layer.

For the travelling wave with the dispersion relationship $F(\omega_0, k, \omega) = 0$, expanding $\omega$ as a Taylor's series expansion and substituting for $\Delta\omega_0$ ($= 1/T_{FMR}$) leads to $\omega(\omega_0 + \Delta\omega_0) = \omega(\omega_0) + [j\alpha\omega_0]\partial_{\omega_0}\omega$. Note that, to the first order, the imaginary part of the expression is responsible for damping of spin waves. The imaginary part may be conveniently evaluated using the implicit function $\alpha\omega_0\partial_{\omega_0}\omega = \alpha\omega_0\partial_{\omega_0}F/\partial_\omega F$. Using the dispersion relation for MSSW and substituting $\omega_{MSSW}$ as $\omega$, the relaxation time for surface waves can be calculated as $1/T_{MSSW} = \alpha(\omega_0 + \omega_M/2)$. Furthermore, the group velocity of MSSW waves is given by $v_{MSSW} = \partial_k\omega = d\omega_M^2 e^{-2kd}/4\omega$. Hence, the distance over which spin wave signal drops by $e^{-1}$ of its original value is $L = v_{MSSW}T_{MSSW}$.

As the bias field increases, so does the frequency. At higher frequencies, the group velocity of the spin waves slows down as $1/\omega$. Hence, the distance that spin waves travel before they attenuate by $e^{-1}$ also reduces by the same factor. This decrease in the effective area over which spin pumping due to spin wave excitation is dominant, effectively cancels the increase in the amplitude of the spin pumping signal with $\omega$ [see the $j\omega m_z$ term in Eq. (3)].

Note that this is an effective area over which the current can be normalized over. Hence, one needs a proportionality constant by which the value of the real current may be extracted from the measured value of spin pumping voltage. Finally, since the spin wave travels in both directions, from both the signal line and ground line, and that the signal under the ground line is weaker than that under the ground line by a factor of 3, the length of the section of the Py responsible for spin pumping due to spin waves is given by:

$$L_{sw} = 2\left(1+\frac{1}{3}\right) A_{sc} \frac{d\omega_M^2 \, e^{-2kd}}{4\omega\alpha\left(\omega_0 + \omega_M/2\right)}. \quad (7)$$

Equations (3), (5), and (7) can be simultaneously be used for finding the spin pumping voltage. The only unknown parameter in the equations is $A_{sc}$. In Fig. 3(a), the measured values have been fitted with $A_{sc} = 20.625$. As can be seen from Fig. 3(a), there is very good agreement between the measurements and calculation. It is interesting to see the value of the length $A_{sc}L$, which is plotted in Fig. 3(b). The effective length is in the range of 100 μm to 200 μm and decreases significantly with the precessional frequency. As Bailleul et al. [28] has measured spin wave signals up to 100 μm using a VNA, the extracted effective length shows that spin waves are able to travel 100 – 200 μm. If we use a spin Hall angle of 0.0067 for the above calculation, the effective length ($A_{sc}L$) will be ~55 times bigger, which becomes 5 – 10 mm and is much larger than the typical propagation length of spin waves in Py.

**V. INFLUENCE OF AMR**

An RF current is able to set up an AMR homodyne dc voltage given by [31]:

$$V_{AMR} = i_{rf}^m \frac{R_{wg}}{R_{Pt\|Py}} \Delta R_{AMR} \frac{\sin(2\theta)\sin(2\alpha)}{4} \cos\varphi_0, \tag{8}$$

where $R_{wg}$ and $R_{Pt\|Py}$ are the resistances of the waveguide and the sample respectively, $\Delta R_{AMR}$ is the difference in the maximum and minimum value of the sample resistance due to AMR, $\theta$ is the precessional cone angle, $\alpha$ is the angle that the direction of the magnetization makes with the waveguide axis. $\varphi_0$ is the phase angle between the precession and the driving rf field. In the configuration of the current measurement $\alpha = 0$. Hence, $V_{AMR} = 0$.

In order to make sure that the influence of AMR is negligible in our case, another device on the same sample, in which the Pt is not present, but the Py is extended beyond the oxide, is measured. The cross section of this device is shown in Fig. 4(a). In this device, any AMR voltage which might be produced can be directly measured. The observed AMR voltage ($V_{AMR}$) was negligible in comparison to the spin pumping voltage ($V_{SP}$). A comparison between the AMR voltage measured across the device shown in Fig. 4(a), and inverse spin Hall voltage measured under the same conditions across the device shown in Fig. 1, is shown in Fig. 4(b). Both devices are on the same wafer and the applied field is 13.1 mT for both measurements. It is clear that there is no AMR contribution in our data.

## VI. FREQUENCY CHARACTERISTIC OF THE SPIN PUMPING SIGNAL

In order to find the frequency characteristics of the spin pumping signal, measurements of the spin pumping signal is performed for many values of magnetic fields. This frequency dependence is shown as a contour plot in Fig. 5. Also plotted on top of the contour plot are the dispersion characteristics of the MSSW mode $\omega_{MSSW}$ [see Eq. (6)] by a dashed line, and $\omega_{FMR} = \omega_{MSSW}(k = 0)$ by a solid line. The frequency characteristics of the measurements shown in Fig. 2 and Fig. 3(a), have been plotted as open squares. It is easily seen that the spin pumping voltage is only observable for magnetization dynamics for frequencies corresponding to those of either MSSW or FMR. We further show from the calculations of the amplitude of the spin pumping signal in the next section that the MSSW dominates the spin pumping process.

## VII. DECAY CHARACTERSITICS OF THE SPIN PUMPING SIGNAL

Due to the significant dependence of $L_{sw}$ on the precessional frequency [see Eq. (7)] of the magnetization, the variation of the amplitude of the spin pumping signals resulting from spin waves will have significantly different characteristics than that arising from FMR. In Fig. 6(a), the amplitude of measured spin pumping voltages shown in Fig. 5 is plotted as filled squares. The amplitude is calculated as the difference between the maximum and the minimum value of the measured spin pumping signal. The amplitude of the signal calculated using Eq. (5) is also plotted by a solid line. As can be seen, the amplitude of the measured signals and the calculated amplitudes match well. The calculated amplitude of the spin pumping signal resulting from FMR only is plotted as a dashed line. As can be seen, the amplitude of the spin pumping signal resulting from the FMR only is significantly smaller than that of that resulting from spin waves.

## VIII. CONCLUSION

In conclusion, we have presented the theoretical foundation for extending the prevalent theory of spin pumping resulting from FMR to include spin pumping resulting from spin waves. A method for determining the area over which the spin pumping process occurs has been demonstrated, and has been shown to be significantly greater than the signal and ground line overlap area. Furthermore, from the calculations of the change in the amplitude of the spin pumping signal as a function of the applied bias field, it is apparent that MSSW dominates the spin pumping process. Hence, it has been shown that MSSW can he used as a very efficient method for increasing the spin pumping signal compared to the FMR only case.

**ACKNOWLEDGEMENT**

This work is supported by the Singapore National Research Foundation under CRP Award No. NRF-CRP 4-2008-06.

*eleyang@nus.edu.sg

# References

[1] L. Berger, Phys. Rev. B **54**, 9353 (1996).
[2] L. Berger, Phys. Rev. B **59**, 11465 (1999).
[3] T. Taniguchi and H. Imamura, Phys. Rev. B **76**, 092402 (2007).
[4] Y. Tserkovnyak, A. Brataas, and G. E. W. Bauer, Phys. Rev. Lett. **88**, 117601 (2002).
[5] S. Mizukami, Y. Ando, and T. Miyazaki, Phys. Rev. B **66**, 104413 (2002).
[6] R. Urban, G. Woltersdorf, and B. Heinrich, Phys. Rev. Lett. **87**, 217204 (2001).
[7] J. Xiao, G. E. W. Bauer, S. Maekawa, and A. Brataas, Phys. Rev. B **79**, 174415 (2009).
[8] A. Azevedo, L. H. V. Leao, R. L. Rodriguez-Suarez, A. B. Oliveira, and S. M. Rezende, J. Appl. Phys. **97**, 10C715 (2005).
[9] M. V. Costache, M. Sladkov, S. M. Watts, C. H. van der Wal, and B. J. van Wees, Phys. Rev. Lett. **97**, 216603 (2006).
[10] H. Y. Inoue, K. Harii, K. Ando, K. Sasage, and E. Saitoh, J. Appl. Phys. **102**, 083915 (2007).
[11] M. V. Costache, S. M. Watts, C. H. van der Wal, and B. J. van Wees, Phys. Rev. B **78**, 064423 (2008).
[12] K. Ando, S. Takahashi, J. Ieda, Y. Kajiwara, H. Nakayama, T. Yoshino, K. Harii, Y. Fujikawa, M. Matsuo, S. Maekawa, and E. Saitoh, J. Appl. Phys. **109**, 103913 (2011).
[13] M. Gradhand, D. V. Fedorov, P. Zahn, and I. Mertig, Phys. Rev. B **81**, 245109 (2010).
[14] K. Harii, T. An, Y. Kajiwara, K. Ando, H. Nakayama, T. Yoshino, and E. Saitoh, J. Appl. Phys. **109**, 116105 (2011).
[15] O. Mosendz, V. Vlaminck, J. E. Pearson, F. Y. Fradin, G. E. W. Bauer, S. D. Bader, and A. Hoffmann, Phys. Rev. B **82**, 214403 (2010).
[16] O. Mosendz, J. E. Pearson, F. Y. Fradin, S. D. Bader, and A. Hoffmann, Appl. Phys. Lett. **96**, 022502 (2010).
[17] L. H. Vilela-Leao, G. L. da Silva, C. Salvador, S. M. Rezende, and A. Azevedo, J. Appl. Phys. **109**, 07C910 (2011).
[18] A. Azevedo, L. H. Vilela-Leão, R. L. Rodríguez-Suárez, A. F. Lacerda Santos, and S. M. Rezende, Phys. Rev. B **83**, 144402 (2011).
[19] L. H. Vilela-Leao, C. Salvador, A. Azevedo, and S. M. Rezende, Appl. Phys. Lett. **99**, 102505 (2011).
[20] T. Yoshino, K. Ando, K. Harii, H. Nakayama, Y. Kajiwara, and E. Saitoh, Appl. Phys. Lett. **98**, 132503 (2011).
[21] H. Chudo, K. Ando, K. Saito, S. Okayasu, R. Haruki, Y. Sakuraba, H. Yasuoka, K. Takanashi, and E. Saitoh, J. Appl. Phys. **109**, 073915 (2011).
[22] L. Liu, T. Moriyama, D. C. Ralph, and R. A. Buhrman, Phys. Rev. Lett. **106**, 036601 (2011).
[23] K. Ando, T. An, and E. Saitoh, Appl. Phys. Lett. **99**, 092510 (2011).
[24] K. Ando, S. Takahashi, J. Ieda, H. Kurebayashi, T. Trypiniotis, C. H. W. Barnes, S. Maekawa, and E. Saitoh, Nat. Mater. **10**, 655 (2011).
[25] Y. Kajiwara, K. Harii, S. Takahashi, J. Ohe, K. Uchida, M. Mizuguchi, H. Umezawa, H. Kawai, K. Ando, K. Takanashi, S. Maekawa, and E. Saitoh, Nature **464**, 262 (2010).
[26] C. W. Sandweg, Y. Kajiwara, K. Ando, E. Saitoh, and B. Hillebrands, Appl. Phys. Lett. **97**, 252504 (2010).
[27] H. Kurebayashi, O. Dzyapko, V. E. Demidov, D. Fang, A. J. Ferguson, and S. O. Demokritov, Appl. Phys. Lett. **99**, 162502 (2011).
[28] M. Bailleul, D. Olligs, and C. Fermon, Appl. Phys. Lett. **83**, 972 (2003).
[29] T. J. Silva, C. S. Lee, T. M. Crawford, and C. T. Rogers, J. Appl. Phys. **85**, 7849 (1999).

[30] G. Counil, J. Kim, T. Devolder, C. Chappert, K. Shigeto, and Y. Otani, J. Appl. Phys. **95**, 5646 (2004).
[31] O. Mosendz, J. E. Pearson, F. Y. Fradin, G. E. W. Bauer, S. D. Bader, and A. Hoffmann, Phys. Rev. Lett. **104**, 046601 (2010).
[32] K. Ando and E. Saitoh, J. Appl. Phys. **108**, 113925 (2010).
[33] L. Vila, T. Kimura, and Y. Otani, Phys. Rev. Lett. **99**, 226604 (2007).
[34] T. Kimura, Y. Otani, T. Sato, S. Takahashi, and S. Maekawa, Phys. Rev. Lett. **98**, 156601 (2007).
[35] D. D. Stancil, J. Appl. Phys. **59**, 218 (1986).
[36] M. Covington, T. M. Crawford, and G. J. Parker, Phys. Rev. Lett. **89**, 237202 (2002).
[37] K. Sekiguchi, K. Yamada, S. M. Seo, K. J. Lee, D. Chiba, K. Kobayashi, and T. Ono, Appl. Phys. Lett. **97**, 022508 (2010).

**Figure Captions:**

Fig. 1. (a) A schematic representation of the device geometry (not to scale) and the measurement setup for the study of spin pumping and spin waves. A rectangular Py layer is patterned on top of a longer Pt section and is subsequently insulated using sputtered $SiO_2$. ACPS lines are patterned on top of the $SiO_2$. A signal generator (SG) is connected to one ACPS line, while the other is connected to a spectrum analyzer (SA). A voltmeter is connected across the Pt for measuring the spin pumping signal. (b) Cross section of the device. The layer stack is Pt (10)|$Ni_{81}Fe_{19}$ (20)|$SiO_2$ (30) (thickness is in nm). The bottom Pt is extended for making dc electrical contacts.

Fig. 2. The measured (open squares) and calculated (solid lines) values of the spin wave power spectra ($P_{sw}$) as a function of the precessional frequency ($f$) for various values of the applied bias field $H_b$.

Fig. 3. (a) The measured (open squares) and calculated (solid lines) values of the spin pumping voltage ($V_{SP}$) as a function of the precessional frequency ($f$) for various values of the applied bias field $H_b$. (b) The effective length ($A_{sc}L$) used for quantifying the spin pumping signal resulting from spin waves as a function of the precessional frequency.

Fig. 4. (a) A schematic cross-section of the device used for measuring the voltage resulting from a homodyne AMR contribution of the precessional magnetization. It comprises of an ACPS waveguide patterned on top of an insulated NiFe film, with electrical contact leads at the two ends of the NiFe. This device is present on the same wafer as the device shown in Fig. 1, and fabricated simultaneously, and hence all thicknesses and materials are exactly the same. (b) The measured spin pumping signal from the device shown in Fig. 1 is plotted for an applied bias field of $H_b$=13.1 mT as open squares, while the measured AMR voltage from the device shown in Fig. 4(a) is plotted as open circles to show that there is no contribution of AMR in the present setup.

Fig. 5. A filled contour plot of the spin pumping voltage as a function of the applied bias magnetic field ($H_b$) and frequency ($f$). A white solid line shows the calculated FMR frequency as a function of $H_b$. The red dashed line show the calculated surface wave frequencies. The measured frequency resulting from the simultaneous spin wave and spin pumping measurement in Fig. 2 and Fig. 3(a) are shown as open squares.

Fig. 6. The amplitude of measured spin pumping signals shown in Fig. 5 is plotted as a function of the precessional frequency $f$ as solid squares. The amplitude of the calculated spin pumping voltage as a function of $f$ due to both FMR and MSSW is plotted as a solid line. Also plotted by a dashed line is the contribution of the spin pumping signal resulting from FMR only.

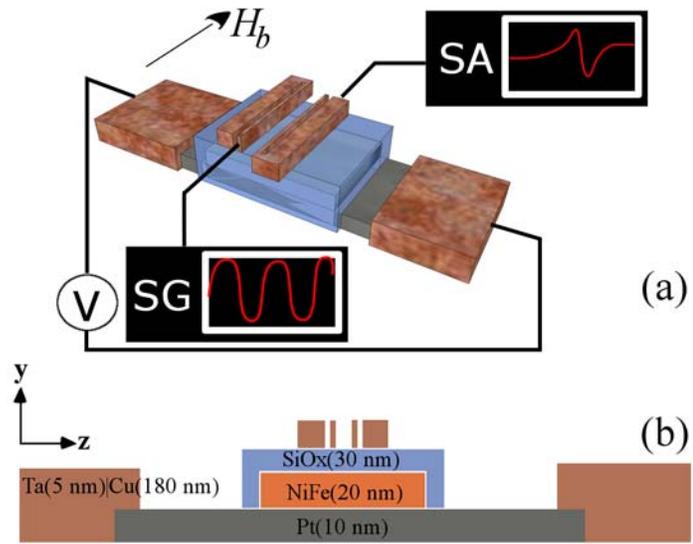

Figure 1

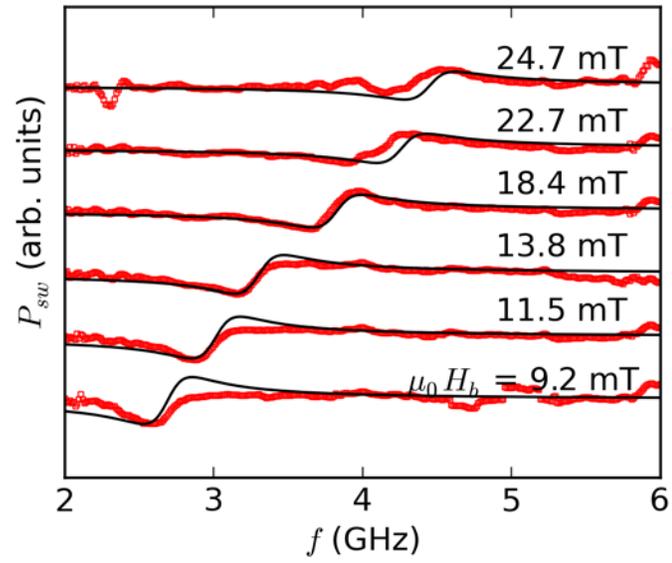

Figure 2

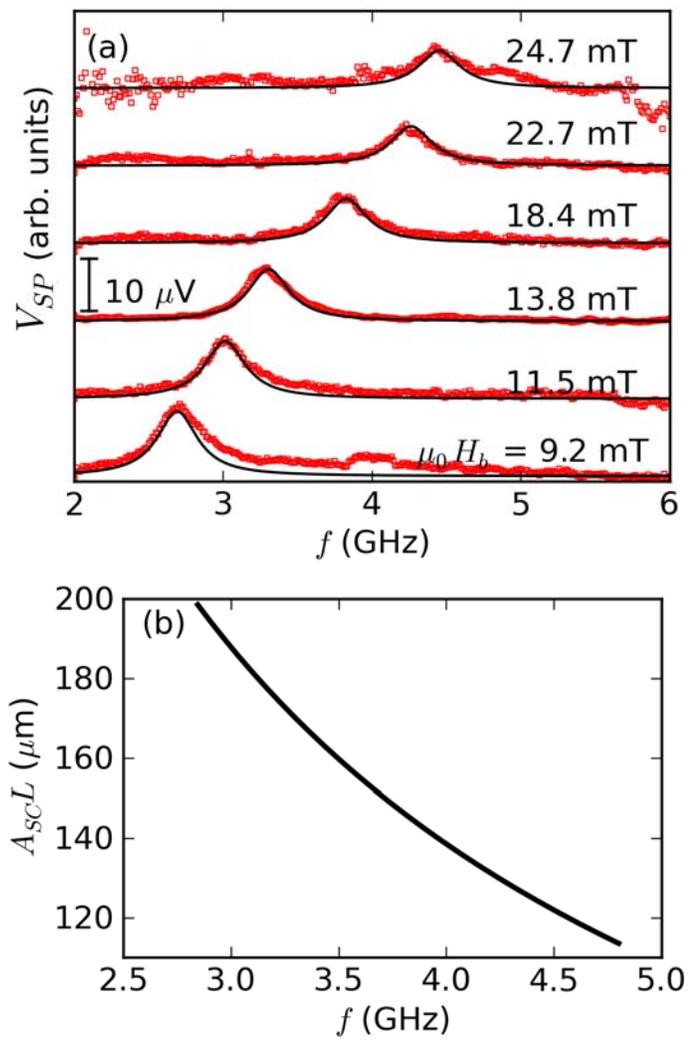

Figure 3

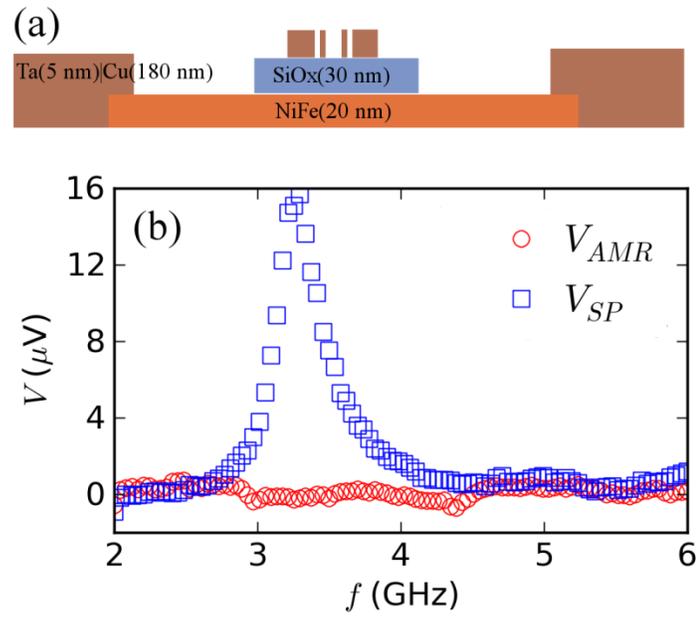

Figure 4.

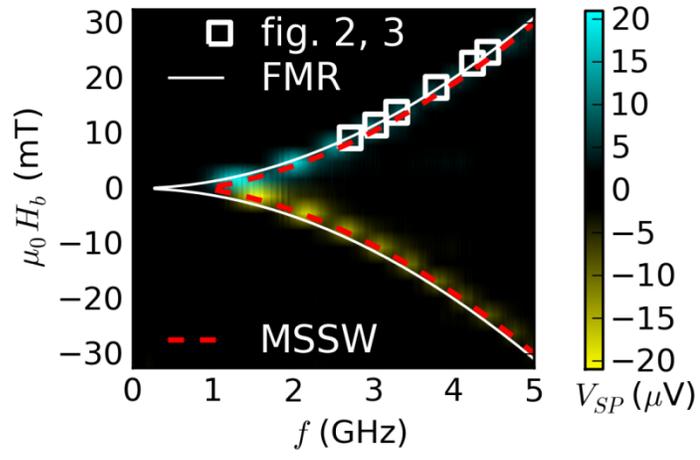

Figure 5

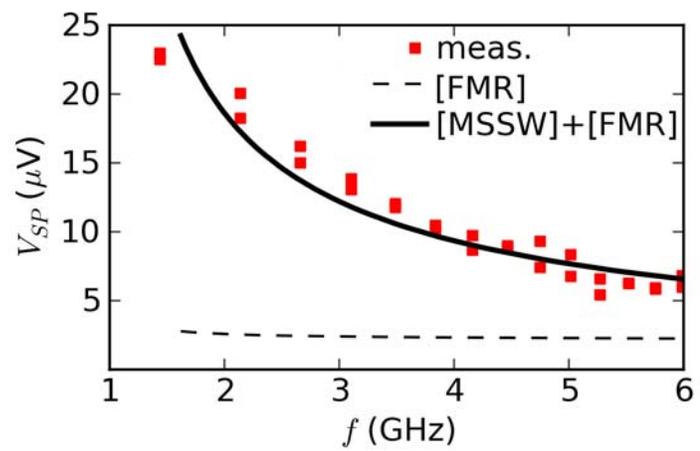

Figure 6.